\documentclass[
    aps,
    prb,
    reprint,
    superscriptaddress
]{revtex4-2}

\usepackage[T1]{fontenc}

\usepackage{float}
\usepackage{amsmath}
\usepackage{amssymb}
\usepackage{graphicx}
\usepackage{setspace}
\usepackage{esint}
\usepackage{color}

\usepackage{svg}
\usepackage[normalem]{ulem}
\usepackage{ragged2e}
\usepackage{placeins}
\usepackage{paralist}
\usepackage{xcolor}

\makeatletter

\@ifundefined{textcolor}{}
{%
    \definecolor{BLACK}{gray}{0}
    \definecolor{WHITE}{gray}{1}
    \definecolor{RED}{rgb}{1,0,0}
    \definecolor{GREEN}{rgb}{0,1,0}
    \definecolor{BLUE}{rgb}{0,0,1}
    \definecolor{CYAN}{cmyk}{1,0,0,0}
    \definecolor{MAGENTA}{cmyk}{0,1,0,0}
    \definecolor{YELLOW}{cmyk}{0,0,1,0}
}

\@ifundefined{textcolor}{}{%
    \definecolor{BLACK}{gray}{0}
    \definecolor{WHITE}{gray}{1}
    \definecolor{RED}{rgb}{1,0,0}
    \definecolor{GREEN}{rgb}{0,1,0}
    \definecolor{BLUE}{rgb}{0,0,1}
    \definecolor{CYAN}{cmyk}{1,0,0,0}
    \definecolor{MAGENTA}{cmyk}{0,1,0,0}
    \definecolor{YELLOW}{cmyk}{0,0,1,0}
}

\def\btt#1{\texttt{\@backslashchar#1}}
\DeclareRobustCommand{\bblash}{\btt{\@backslashchar}}

\@ifundefined{showcaptionsetup}{}{%
    \PassOptionsToPackage{caption=false}{subfig}
}

\usepackage{subfig}

\makeatother

\usepackage{babel}

\usepackage[
    colorlinks=true,
    linkcolor=blue,
    citecolor=black,
    urlcolor=magenta
]{hyperref}

\graphicspath{{images/}}

\begin{document}

\title{Correlated Low-Energy Model of Monolayer 1H-NbS$_2$:
A cRPA+DMFT Study}
\author{Karollaine C. Leite}
\email{karollaine@fisica.ufmt.br}
\affiliation{Instituto de F\'isica, Universidade Federal de Mato Grosso,
78060-900, Cuiab\'a, Mato Grosso, Brazil.}

\author{Jorge L. B. de Faria}
\affiliation{Faculdade de Engenharia, Universidade Federal de Mato Grosso,
78060-900, Cuiab\'a, Mato Grosso, Brazil.}

\author{Luiz Antonio Ribeiro Junior}
\affiliation{Department of Physics, Norwegian University of Science and Technology (NTNU),
NO-7491 Trondheim, Norway}
\affiliation{Computational Materials Laboratory, LCCMat, Institute of Physics,
University of Bras{\'{i}}lia, 70910-900, Bras{\'{i}}lia, Brazil}

\author{Alexandre C. Dias}
\affiliation{Institute of Physics and International Center of Physics,
University of Bras{\'{i}}lia, Bras{\'{i}}lia, 70910-900, Brazil}
\affiliation{Computational Materials Laboratory, LCCMat, Institute of Physics,
University of Bras{\'{i}}lia, 70910-900, Bras{\'{i}}lia, Brazil}

\author{Helena Bragança}
\affiliation{Institute of Physics and International Center of Physics,
University of Bras{\'{i}}lia, Bras{\'{i}}lia, 70910-900, Brazil}

\author{Alberto S. de Arruda}
\email{aarruda@fisica.ufmt.br}
\affiliation{Instituto de F\'isica, Universidade Federal de Mato Grosso,
78060-900, Cuiab\'a, Mato Grosso, Brazil.}



\begin{abstract}

Monolayer $1\mathrm{H}\text{-}\mathrm{NbS}_2$ hosts a well-separated metallic band of predominant Nb-$d_{z^2}$ character, providing an ideal platform to probe electronic correlations within a single-band setting. In this work, we present a comprehensive study of $1\mathrm{H}\text{-}\mathrm{NbS}_2$ by combining first-principles Wannier interpolation, constrained random-phase approximation (cRPA), and DFT+DMFT. We show that an effective single-band model accurately captures the low-energy electronic structure, reproducing key experimental features from ARPES and STS $dI/dV$ spectra, including a characteristic Van Hove singularity. While static DFT+$U$ fails to describe the correlated metallic nature of the system, DFT+DMFT successfully accounts for the spectral weight redistribution and dynamic correlations. By defining the cRPA target subspace directly within the isolated Nb-$d_{z^2}$ band, we obtain a renormalized local interaction of $U_{\mathrm{cRPA}} = 1.138~\mathrm{eV}$. This interaction stabilizes a strongly correlated metallic state that remains robust upon cooling. Remarkably, this minimal single-band model captures a spectral broadening similar to that previously reported in more elaborate treatments - incorporating larger local interactions ($U \sim 1.8~\mathrm{eV}$, derived from a multiorbital cRPA construction), intersite Coulomb terms ($V$), and electron-phonon coupling - demonstrating that a consistently downfolded local interaction alone can already capture key aspects of the experimentally observed low-energy behavior of monolayer $1\mathrm{H}\text{-}\mathrm{NbS}_2$.

\end{abstract}
\maketitle


\section{INTRODUCTION}
Since graphene was first isolated \cite{geim1,neto,geim2}, transition metal dichalcogenides (TMDs), which are materials with the general formula $MX_2$, where M stands for a transition metal and X for a chalcogen (S, Se, or Te), have established themselves as one of the most promising families of two-dimensional materials. Unlike graphene, however, TMDs truly stand out for their structural and electronic richness, since they can crystallize into a variety of polymorphic phases, including the 1T, 1H, and 3R structures \cite{Liu,zhai,guller,mani,kuma,kis,afaf,jasil}, giving this class a versatility rarely seen in other 2D systems.

Among the TMD family, the metallic ones from group V, like
$NbS_2, NbSe_2$, $TaS_2$, and $TaSe_2$, deserve special
attention \cite{kis}.  What makes them so interesting is the way their layered structure pairs with genuinely metallic behavior. That metallicity comes down to the partially filled $d$-orbitals of the transition metal, which keep the material highly conductive even when you thin it down to a single monolayer \cite{qu}. Niobium disulfide ($NbS_2$), in particular, has become something of a go to system for probing electronic correlations and collective effects in two dimensions \cite{Kawa,du}. Over the past couple of decades, a growing number of experimental and theoretical work has revealed a much richer picture of its electronic behavior \cite{Lian2023}, for instance, evidence points to two gap superconductivity, orbital-dependent coupling between electrons and phonons, and even a hint of an inclination toward charge density wave (CDW) ordering, although this CDW instability appears to be contained by anharmonic lattice effects \cite{youbi,hell,alshemi}.
Upon reducing this material to the two-dimensional limit, these effects can become considerably more pronounced, as reduced dimensionality enhances the relevance of electronic interactions making everything from superconducting gaps to phonon anomalies harder to ignore \cite{Knispel}.
Take the $1H‑NbS_2$ monolayer: it adopts a trigonal prismatic coordination, with the niobium atoms at the center of sulfur prisms that tile the two-dimensional plane. This coordination geometry is a defining hallmark of the 1H phase, observed in various two-dimensional TMDs \cite{kis}, and gives the monolayer a structural backbone similar to $MoS_2$, albeit with a distinctly different electronic configuration.

The electronic structure of the $1H-NbS_2$ monolayer continues to attract considerable interest. Angle‑resolved photoemission spectroscopy (ARPES) conducted on epitaxial monolayers grown on Au(111) corroborates the existence of a metallic band structure that remains stable down to at least 30 K \cite{Stan}. In addition, more recent ARPES investigations reveal that the Fermi surface of the $1H-NbS_2$ monolayer comprises two primary hole pockets, centered at the high symmetry $\Gamma$ and $K$ points of the Brillouin zone \cite{Kawa,youbi}. The morphology of these pockets is crucial in dictating the low‑energy electronic behavior of the system, as it directly influences its propensity toward collective ground states, including charge density waves (CDWs) and superconductivity.

A particularly noticeable feature of the ARPES spectra is the pronounced broadening of the electronic bands. Two main ideas have been proposed to explain this phenomenon. The first attributes the cause to the underlying Au(111) substrate, that is, the monolayer hybridizes with the metallic support, which effectively dampens and broadens the electronic states \cite{Stan}. The other group, however, argues that the broadening is primarily intrinsic, rooted in the many-body physics that the material already exhibits. In this regard, work by van Loon and collaborators \cite{Loon} demonstrated that, when local and non-local Coulomb interactions are included alongside electron-phonon coupling, the resulting spectrum appears strongly renormalized and noticeably broadened, exactly the type of line profile observed experimentally.

Despite these advances, an important question concerns the effective interaction appropriate for a minimal single-band description of monolayer 1H-NbS$_2$. In Ref.~\cite{Loon}, the partially screened Coulomb interaction was evaluated in a three-orbital Nb-$d$ basis ($d_{z^2}$, $d_{xy}$, and $d_{x^2-y^2}$) and subsequently reduced to a single-band interaction by neglecting the orbital dependence and averaging the matrix elements of the multi-orbital interaction. This procedure yielded an effective on-site interaction of approximately $U = 1.8$~eV for a metallic band with a bandwidth of about $W = 1.2$~eV. When considered as the local Coulomb interaction alone, this interaction scale was found to drive the half-filled single-band model into an insulating regime, whereas the inclusion of nonlocal Coulomb and electron-phonon interactions restored a correlated metallic state through their competing effects \cite{Loon}. This approach, while comprehensive in its inclusion of competing
interactions, leaves open the question of whether a subspace-consistent local interaction alone - derived directly within the single-band low-energy subspace- is sufficient to stabilize a correlated metallic state in the minimal model. It is critical to note that cRPA interactions depend on the definition of the low-energy subspace and on the screening channels excluded in the construction of the effective interaction. 

Recent systematic cRPA studies of $MX_2$ monolayers report a
single-orbital $U$ value for $1H$-NbS$_2$ of $1.29$~eV
\cite{Ramezani}, in close agreement with the value obtained here.
Furthermore, comparative studies of cRPA and linear response theory
have shown that the cRPA scheme can yield unrealistically small $U$
values when the target and screening subspaces hybridize strongly
\cite{carta2026}, underscoring the importance of a consistent subspace
definition.
Complementary information on the local electronic structure is provided by scanning tunneling spectroscopy (STS) measurements of monolayer 1H-NbS$_2$ on graphene/Ir(111)~\cite{Knispel}. The measured $dI/dV$ spectrum exhibits a pronounced feature near $-0.15$~eV, identified as a Van Hove singularity associated with the toroidal minimum of the Nb-derived $d$ band around the $\Gamma$ point. This feature provides a particularly useful experimental reference for the Nb-derived low-energy band represented in the single-band construction of this work. In contrast, spectral structures at larger binding energies, such as the feature near $-1.25$~eV attributed to occupied S-$p$ states, lie outside the Nb-$d_{z^2}$ subspace retained in our effective model.
Motivated by these considerations, we construct an effective single-band
model for the isolated band of predominant Nb-$d_{z^2}$ character,
determine the local Coulomb interaction directly within the same
low-energy subspace, and investigate its many-body consequences within finite-temperature DMFT.

The outline of this paper is as follows: In Section II, we present the theoretical and computational methodology. This includes the low-energy model for monolayer $1H$-NbS$_2$, the DFT calculations, the Wannier downfolding procedure, and the determination of the effective Coulomb interaction within cRPA. We further describe the DFT+$U$ and DFT+DMFT frameworks, detailing the CT-HYB impurity solver, the interaction scan, and the temperature-dependent analysis of the single-band model. In Section III, we present the results and discussion, covering the DFT band structure, the validation of the single-band Wannier model against DFT and STS data, the correlation-driven evolution of the spectral function, and the analysis of the self-energy, quasiparticle weight, and effective mass enhancement. Finally, Section IV summarizes our main conclusions.

\begin{figure}[ht]
    \centering
     \includegraphics[width=0.8\linewidth]{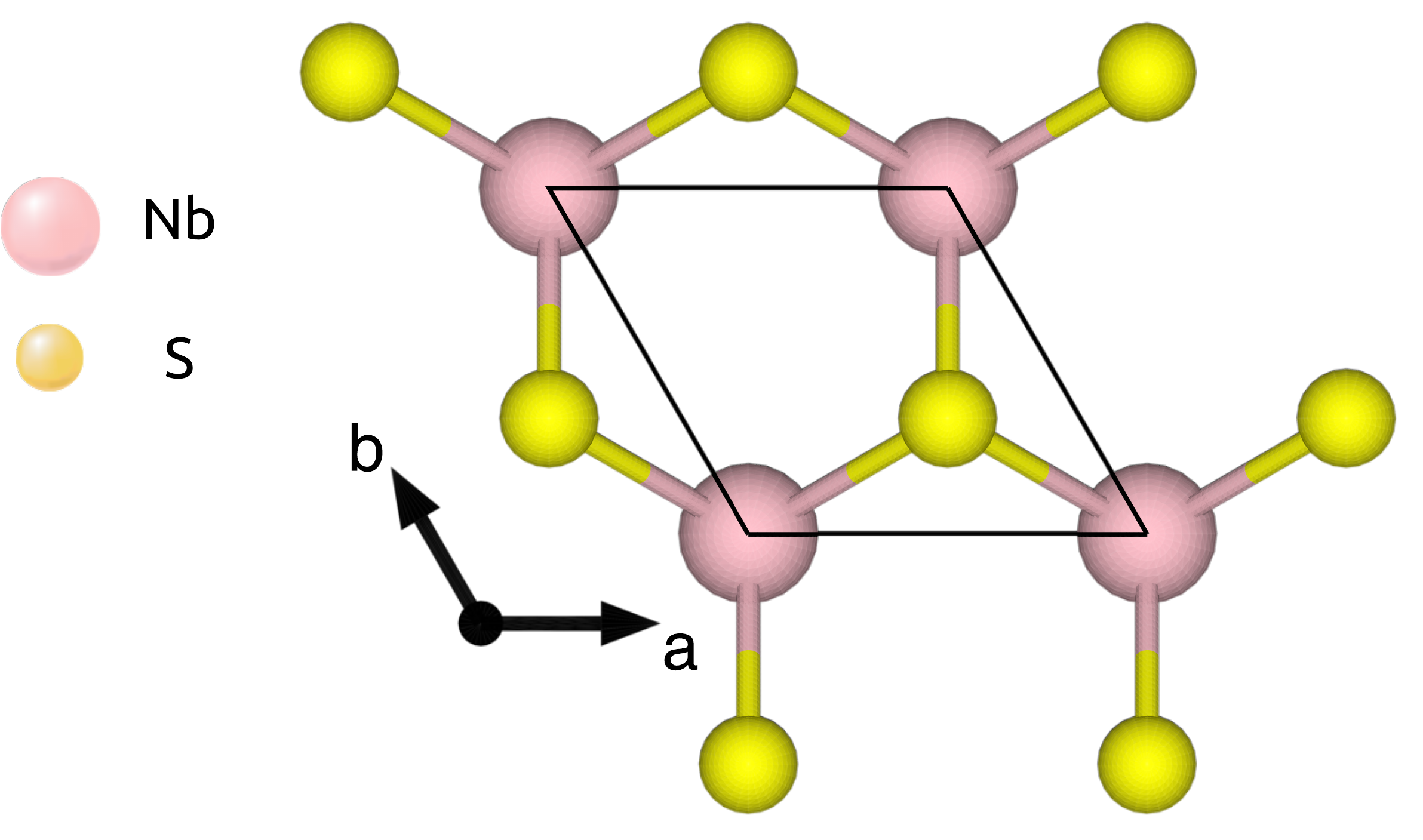}
    \caption{\footnotesize\justifying
    Crystal structure of monolayer NbS$_2$ in the 1H phase.
    Each Nb atom is coordinated by six S atoms in a trigonal-prismatic
    geometry, forming a two-dimensional S--Nb--S layer.}
    \label{crystal_structure}
\end{figure}


\section{THEORETICAL AND COMPUTATIONAL DETAILS}

In this section, we present the theoretical and computational framework
employed in this work. We first introduce the low-energy model for
monolayer $1H$-NbS$_2$, followed by the DFT calculations and the
Wannier downfolding used to construct the single-band low-energy
Hamiltonian. We then describe the determination of the local Hubbard
interaction within cRPA, the treatment of static local correlations
within DFT+$U$, and, finally, the DFT+DMFT calculations used to
investigate dynamical correlation effects.

\subsection{The Low-energy model and calculation}

The low-energy electronic properties of monolayer 1H-NbS$_2$ were described here by a single-band Hubbard model \cite{Hubbard1963}.

The effective Hamiltonian is given by
\begin{equation}
\label{hamiltonian}
H=
\sum_{\langle ij\rangle,\sigma}
t_{ij}
c^{\dagger}_{i\sigma}
c_{j\sigma}
+
U\sum_i
n_{i\uparrow}
n_{i\downarrow}
-
\mu
\sum_{i,\sigma}
c^{\dagger}_{i\sigma}
c_{i\sigma}.
\end{equation}
where $t_{ij}$ denotes the hopping amplitudes between Wannier orbitals located at sites $i$ and $j$, which were extracted from the Wannier downfolding of the DFT band structure. The operators $c^{\dagger}_{i\sigma}$ ($c_{i\sigma}$) creates (annihilates) an electron with spin $\sigma$ at site $i$, $n_{i\sigma}=c^{\dagger}_{i\sigma}c_{i\sigma}$ is the corresponding number operator, $U$ is the local Coulomb interaction, and $\mu$ is the chemical potential.

Electronic correlations were treated within Dynamical Mean-Field Theory (DMFT), where the lattice problem is mapped onto an effective quantum impurity embedded in a self consistent electronic bath \cite{Anderson1961}. Within DMFT, the self energy is assumed to be local but frequency dependent, $\Sigma(\omega)$, allowing the description of dynamical correlation effects beyond conventional band theory.

The interacting lattice Green's function is written as
\begin{equation}
G(\mathbf{k},\omega)
=
\left[
\omega+\mu
-\varepsilon(\mathbf{k})
-
\Sigma(\omega)
\right]^{-1},
\end{equation}
where $\varepsilon(\mathbf{k})$ is the non-interacting dispersion obtained from the Hamiltonian (\ref{hamiltonian}) and $\Sigma(\omega)$ is the local DMFT self-energy.

The momentum resolved spectral function was obtained from the interacting Green's function according to
\begin{equation}
A(\mathbf{k},\omega)
=
-\frac{1}{\pi}
\mathrm{Im}
\,G(\mathbf{k},\omega).
\label{eq:spectral_function}
\end{equation}
This quantity provides direct information about the renormalized electronic dispersion and allows a qualitative comparison with angle resolved photoemission spectroscopy (ARPES) experiments.

In this work, the degree of electronic correlation was also quantified through the quasiparticle weight, 
\begin{equation}
Z=
\left[
1-
\left.
\frac{\partial \mathrm{Im}\Sigma(i\nu_n)}
{\partial \nu_n}
\right|_{\nu_n \rightarrow 0}
\right]^{-1},
\end{equation}
where $\nu_n=(2n+1)\pi/\beta$ are the fermionic Matsubara frequencies, $\beta=1/T$ is the inverse temperature (setting $k_B=1$), and $\Sigma(i\nu_n)$ is the local DMFT self-energy on the imaginary-frequency axis. The quasiparticle weight $Z$ provides a measure of correlation-induced quasiparticle renormalization. In the non-interacting limit, $Z=1$, whereas $Z\rightarrow0$ signals the suppression of coherent quasiparticle excitations and the approach to a Mott insulating state. In practice, the low-frequency derivative entering Eq.~(4) was estimated from the two Matsubara frequencies closest
to zero, $\nu_n=\pm\pi/\beta$, providing a symmetric finite difference
estimate of the slope of $\mathrm{Im}\,\Sigma(i\nu_n)$ around the origin. 

The effective mass enhancement was obtained from
\begin{equation}
\frac{m^*}{m}
=
\frac{1}{Z}.
\end{equation}
Thus, a reduction of $Z$ is directly associated with an increase of the effective mass. The simultaneous suppression of $Z$ and enhancement of $m^*/m$ provide a quantitative measure of the increasing strength of electronic correlations and the progressive loss of quasiparticle coherence.

\subsection{DFT calculations }

The electronic structure of monolayer 1H-NbS$_2$ was obtained within Density Functional Theory (DFT) using the \texttt{QUANTUM ESPRESSO (QE)} package \cite{Giannozzi2009}. Exchange-correlation effects were described within the generalized gradient approximation (GGA) using the Perdew-Burke-Ernzerhof (PBE) functional \cite{Perdew1996}. The crystal structure (see Fig \ref{crystal_structure}.) consists of a trigonal prismatic Nb atom sandwiched between two sulfur layers. A vacuum region of approximately 18~\text{\AA}  was introduced along the out-of-plane direction in order to eliminate spurious interactions between periodic images. The lattice parameter was obtained from structural relaxation and found to be $a = 3.34$~\text{\AA} .

Core-valence interactions were treated using projector augmented wave (PAW) \cite {Blochl1994} and ultrasoft pseudopotentials for Nb and S, respectively. The kinetic-energy cutoffs for the wavefunctions and charge density were set to 80 Ry and 560 Ry. Metallic occupations were treated using the Marzari-Vanderbilt smearing scheme \cite{Marzari1999}. Self-consistent calculations were performed using a Monkhorst-Pack $15\times15\times1$ k-point mesh, while a denser $36\times36\times1$ mesh was employed for the non-self-consistent calculations used in the construction of the low-energy model. The electronic self-consistency cycle was converged using a threshold of $10^{-6}$ Ry.The low-energy calculations were performed without spin-orbit coupling. Additional PBE+SOC calculations confirm that the isolated metallic band
and its overall dispersion are preserved.


\subsection{Downfolding and Low-Energy Hamiltonian}

To construct a low-energy description of monolayer $1H$-NbS$_2$, the
DFT electronic structure was downfolded onto a localized Wannier basis
using the \textsc{wannier90} code \cite{Mostofi2014}. The Wannier
representation provides a direct connection between the
first-principles band structure and an effective tight-binding
Hamiltonian while preserving the orbital character of the low-energy
states \cite{Marzari1997,Mostofi2007,Mostofi2014}.

We therefore
construct a single-band model for this low-energy sector, consistent with
Wannier-based descriptions of the $1H$-NbX$_2$ family and related
transition-metal dichalcogenides
\cite{Heil2018}. The target subspace was represented by
a single Wannier orbital initialized from an Nb-$d_{z^2}$ projector. A disentanglement window extending from $-1.5$ to $+1.5$~eV relative
to the Fermi level was employed in the Wannier construction. The resulting one-band
Wannier Hamiltonian reproduces the DFT dispersion of the metallic band
in the proximity of the Fermi level, providing the noninteracting
low-energy Hamiltonian used in the subsequent many-body calculations.

In real space, this Hamiltonian can be written as
\begin{equation}
H_0=\sum_{ij} t_{ij}\,c_i^\dagger c_j,
\end{equation}
where $t_{ij}$ denotes the hopping amplitude between Wannier orbitals
at sites $i$ and $j$. These matrix elements define the
first-principles tight-binding representation of the low-energy
metallic band and constitute the noninteracting part of the
single-band Hubbard model subsequently treated within DMFT.

The single-band construction is intended to describe the low-energy
sector rather than the complete electronic structure of NbS$_2$.
Its validity in the present context follows from the isolation of the
metallic band near the Fermi level and its predominant
Nb-$d_{z^2}$ character. The quality of this low-energy representation
will be demonstrated in Sec.~III\,B through a direct comparison between
the DFT and Wannier-interpolated band structures. Higher-energy states
are integrated out in the downfolding procedure, while their screening
contribution to the effective local interaction is accounted for
separately within the cRPA construction discussed below.

\subsection{Computing the Local Hubbard Parameter}

The next step in constructing the low-energy interacting model is to
determine the local Hubbard interaction associated with the correlated
subspace defined above. We compute this interaction from first
principles within the constrained random-phase approximation (cRPA)
\cite{Aryasetiawan2004,Vaugier2012,carta2026,Si2025}.

The cRPA target space was defined by the isolated metallic band of
predominant Nb-$d_{z^2}$ character, consistently with the single-band
Wannier model employed in the subsequent DMFT calculations. In the
Kohn--Sham band ordering used here, this target corresponds to band 13.
Accordingly, bands 1--12 and 14--128 were excluded from the target
space in the cRPA calculation, leaving the isolated metallic band as
the correlated subspace.

The electronic response was evaluated using 128 Kohn--Sham states, a
$12\times12\times1$ $k$-point mesh, and a plane-wave energy cutoff of
619.276~eV. The interaction was evaluated in the static limit,
$\omega=0$, consistently with the static Hubbard Hamiltonian used in
the DMFT calculations. For the resulting single-orbital correlated
subspace, we obtain
\begin{equation}
U_{\mathrm{cRPA}}
=
U_{1111}(\omega=0)
=
1.138~\mathrm{eV},
\end{equation}
compared with the corresponding bare Coulomb matrix element
$U_{\mathrm{bare}}=4.7655$~eV.

The reduction from $U_{\mathrm{bare}}$ to $U_{\mathrm{cRPA}}$ reflects
the screening channels outside the target band. Importantly,
$U_{\mathrm{cRPA}}=1.138$~eV is specific to the single-band correlated
subspace adopted here and should therefore be interpreted together
with the corresponding Wannier representation
\cite{Reddy2025,carta2026}. This subspace consistency allows the
cRPA-derived interaction to be used directly in the subsequent
single-band DMFT calculations.

\subsection{Static Local Correlations Within DFT+$U$}

As a first step beyond conventional DFT, we investigated the effect of
local Coulomb interactions on the electronic structure of monolayer
$1H$-NbS$_2$ within DFT+$U$. In this approach, a Hubbard correction is
introduced for the localized Nb-$d$ states, allowing static local
correlation effects to be included at the mean-field level
\cite{Dudarev1998,Himmetoglu2014,Pavarini2021}. We employed the simplified rotationally invariant formulation of
Dudarev \cite{Dudarev1998}, in which the correction is
controlled by the effective interaction

\begin{equation}
U_{\mathrm{eff}} = U-J .
\end{equation}

Since $J=0$ in all DFT+$U$ calculations, $U_{\mathrm{eff}}$ coincides
with $U$. The Hubbard correction was applied to the Nb-$d$ states,
which dominate the metallic band near the Fermi level. We considered \(U=0.5\), \(1.138\), \(1.5\), and \(1.8\) eV, covering the range of interaction strengths considered in the subsequent single-band analysis. Because the DFT+\(U\) correction is applied to the Nb-\(d\) manifold, these calculations are used as a qualitative static reference for the evolution of the electronic structure with interaction strength. Non-spin-polarized calculations were
first performed to determine the effect of the static Hubbard correction
on the electronic dispersion without allowing spin symmetry breaking.
We observed no qualitative changes in the isolated metallic band for
the interaction strengths considered. We therefore allowed spin polarization to examine whether the local interaction could qualitatively modify the electronic structure, including the possible opening of a gap, and stabilize a spin-polarized solution. 

Within DFT+$U$, the Hubbard interaction is treated at the level of a
static mean field. In the absence of static local spin or orbital
polarization, the correction reduces essentially to a level shift
within the correlated manifold. Consequently, if the underlying
Kohn--Sham Hamiltonian is metallic and the symmetry and primitive cell
are preserved, the DFT+$U$ solution remains metallic
\cite{Pavarini2021}. On the other hand, an insulating solution obtained within
DFT+$U$ is associated with static symmetry breaking and is therefore
of Slater rather than genuine Mott character \cite{Pavarini2021}.
Dynamical effects such as frequency-dependent renormalization,
spectral-weight redistribution, and the progressive loss of
low-energy quasiparticle character lie beyond this static description
and are addressed within DMFT \cite{Georges1996,Kotliar2006}.

\subsection{DFT+DMFT Calculations}

Electronic correlations were incorporated within the framework of
Dynamical Mean-Field Theory (DMFT) \cite{Georges1996}, as implemented
in the \texttt{TRIQS/solid\_dmft} package \cite{Merkel2022}.
From DMFT, the interacting lattice problem is mapped onto an
effective quantum impurity model embedded in a self-consistently
determined electronic bath. The resulting frequency dependent
self energy provides access to dynamical correlation effects beyond
static mean-field approaches, including quasiparticle renormalization,
spectral weight redistribution, and the progressive loss of
quasiparticle coherence.

The low-energy one band Hamiltonian obtained from the Wannier
downfolding procedure was supplemented by a local Hubbard interaction
term,

\begin{equation}
H = H_0 + U \sum_i n_{i\uparrow}n_{i\downarrow},
\end{equation}

where $H_0$ is the Wannier Hamiltonian and $U$ denotes the on-site
Coulomb interaction. The resulting auxiliary quantum impurity problem was solved using
the continuous-time hybridization-expansion quantum Monte Carlo
(CT-HYB) solver \cite{Hafermann2013,Hafermann2014}.

In this work, the interaction dependence was investigated at
$\beta=40~\mathrm{eV}^{-1}$ ($T\simeq290$~K) for
$U=0.5$, $1.138$, $1.5$, and $1.8$~eV. The value
$U_{\mathrm{cRPA}}=1.138$~eV corresponds to the single-band cRPA
interaction obtained in the present work. In addition, the temperature
dependence at $U_{\mathrm{cRPA}}$ was investigated over the range
$T\simeq116$--$290$~K to examine the evolution of the correlated
metallic state upon cooling.
Since the present work employs a half-filled single band model, the static Hartree contribution can be absorbed into the chemical potential. Consequently, no explicit double-counting correction was applied. Furthermore, the Hund's coupling was set to $J=0$, consistent with the single-orbital nature of the low-energy model. The DMFT self-consistency cycle was iterated until convergence of the local Green's function and self energy was achieved. From the converged solutions, we computed the local density of states (LDOS), momentum resolved spectral function, Matsubara self energy, quasiparticle weight, and effective mass renormalization. Since the DMFT calculations are carried out on the imaginary frequency axis, an analytic continuation is required to access real frequency observables. To this end, we employed the  Maximum Entropy (MaxEnt) \cite{Jarrell1996} method as implemented in the \texttt{TRIQS} framework. The continuation was performed using the MaxEnt interface provided within the \texttt{solid\_dmft} package, which reconstructs the real frequency spectral function from the imaginary time Green's function obtained in the CT-HYB calculations.

The calculations were performed within a one-shot DFT+DMFT framework. The occupation of the correlated Wannier orbital remained close to half-filling ($n \approx 1$) for all interaction strengths considered, with only small variations throughout the DMFT iterations.

\section{Results and discussion }

We now present the electronic structure and validate the low-energy
model, followed by an analysis of static and dynamical correlation
effects.

\subsection{Electronic Structure of Monolayer 1H-NbS$_2$}

\begin{figure}[ht]
    \centering
    \includegraphics[width=1.0\linewidth]{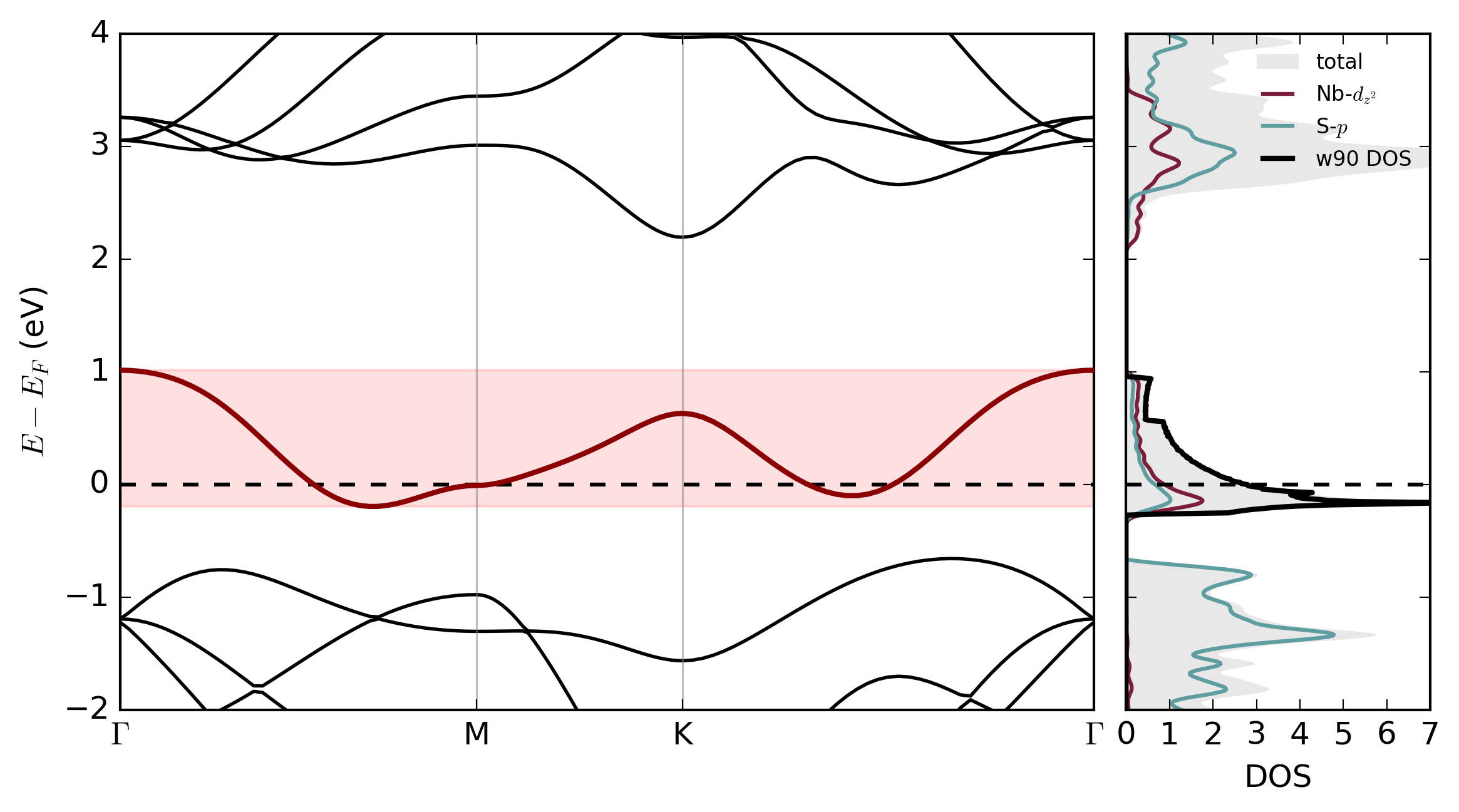}
    \caption{\footnotesize \justifying
DFT band structure (left panel) and projected density of states
(PDOS, right panel) of monolayer $1H$-NbS$_2$. In the left panel,
the vertical axis represents $E-E_F$ (in eV), while the horizontal
axis corresponds to the high-symmetry path
$\Gamma$--M--K--$\Gamma$ of the Brillouin zone; the red line
highlights the band crossing the Fermi level. In the right panel,
the horizontal axis represents the density of states (in states/eV),
and the vertical axis represents $E-E_F$ (in eV). The contributions
from the Nb-$d_{z^2}$ and S-$p$ orbitals, the total DOS, and the DOS
obtained from the single-band Wannier model (black dashed curve) are
shown.
}
    \label{estruturadebanda}
\end{figure}

Figure~\ref{estruturadebanda} shows the DFT band structure and projected
density of states (PDOS) of monolayer 1H-NbS$_2$. A single partially
filled band crosses the Fermi level and is well separated from the
remaining bands over most of the Brillouin zone. The PDOS confirms
that the states near $E_F$ are predominantly derived from the
Nb-$d_{z^2}$ orbital, whereas the S-$p$ contribution becomes more
pronounced away from the Fermi level. The isolated metallic band,
highlighted in red, has a bandwidth of approximately $W=1.2$~eV.

This electronic structure motivates the construction of a single-band
low-energy model centered on the Nb-$d_{z^2}$ orbital. Moreover, since
the interaction strengths considered below are comparable to the
bandwidth of the isolated metallic band, corresponding to $U/W$ ratios
of order unity, correlation effects are expected to play an important
role in the low-energy electronic properties.

\subsection{Low-Energy Electronic Structure}

\begin{figure}[htbp]
    \centering
    \includegraphics[width=0.5\textwidth]{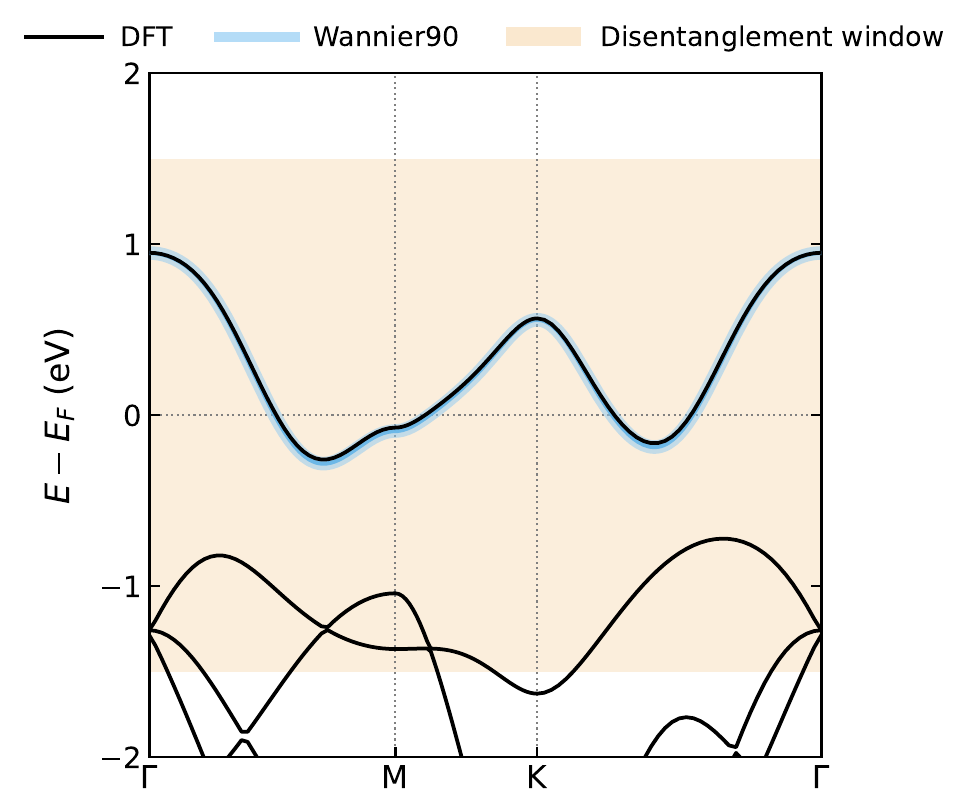}
    \caption{\footnotesize \justifying DFT band structure (black lines) compared with the Wannier90
dispersion (light blue) along the high-symmetry path
$\Gamma$--M--K--$\Gamma$ of the Brillouin zone for monolayer
$1H$-NbS$_2$. The shaded region (light orange) denotes the
disentanglement window used in the downfolding procedure as
implemented in Wannier90. Energies are referenced to the Fermi level.}
    \label{wannier_interpolation}
\end{figure}

The single-band Wannier Hamiltonian accurately reproduces the low-energy DFT dispersion, as shown in Fig.~\ref{wannier_interpolation}. The Wannier-interpolated band closely follows the corresponding DFT band throughout the energy range relevant to the low-energy model, including the Fermi-level crossings. The shaded region indicates the disentanglement window employed in the Wannier construction. The close agreement confirms that the single-band representation captures the low-energy Nb-$d_{z^2}$ sector considered here, providing the electronic basis for the subsequent cRPA+DMFT calculations

\begin{figure}[htbp]
    \centering
    \includegraphics[width=0.5\textwidth]{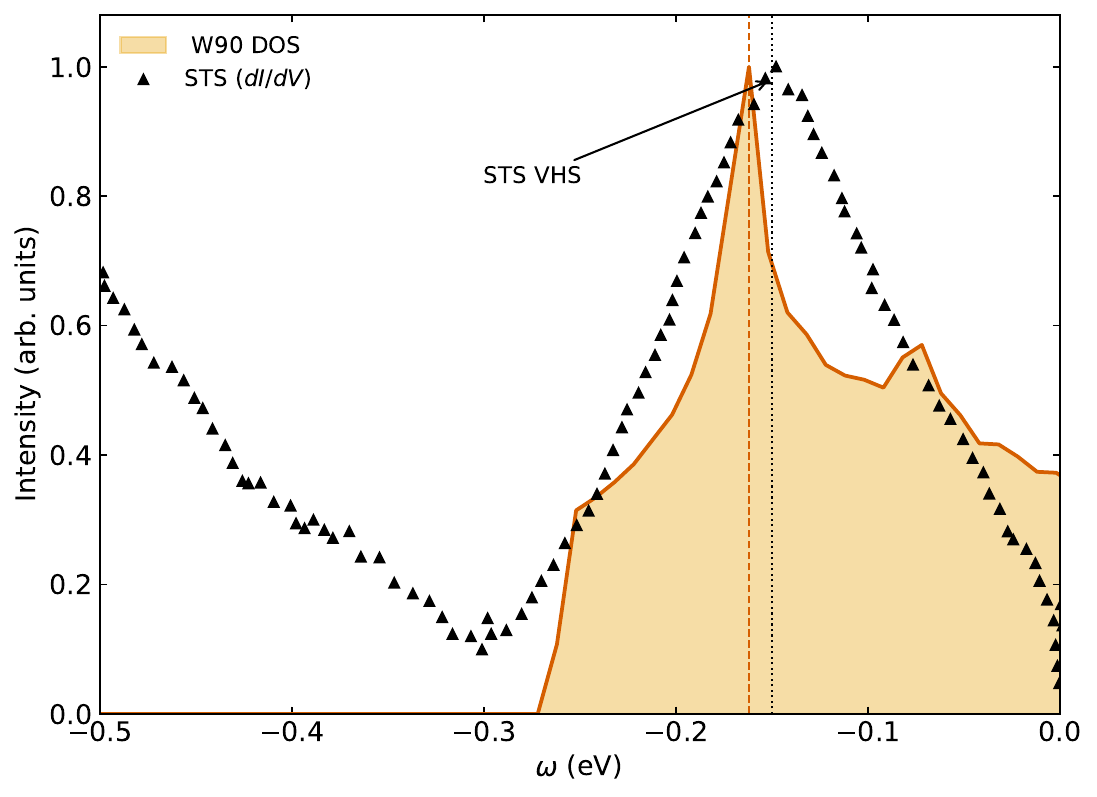}
    \caption{\footnotesize \justifying Comparison between the theoretical density of states obtained from the
single-band Wannier model (W90-DOS, orange filled area) and the
experimental STS $dI/dV$ spectrum of monolayer $1H$-NbS$_2$ on
graphene/Ir(111) at $T_s=0.4$~K from Ref.~\cite{Knispel}.
The vertical axis represents the normalized intensity (arb.~units),
and the horizontal axis represents the energy $E-E_F$ (in eV).
The black dotted and orange dashed vertical lines indicate the
experimental VHS position and the corresponding maximum of the
single-band W90 DOS, respectively.}
    \label{fig:sts_w90}
\end{figure}

The single-band Hamiltonian also reproduces the characteristic
energy scale observed experimentally by scanning tunneling
spectroscopy (STS) for monolayer $1H$-NbS$_2$ on graphene/Ir(111)
at $T_s=0.4$~K \cite{Knispel}. The experimental $dI/dV$ spectrum
shown in Fig.~\ref{fig:sts_w90} was digitized from Fig.~S2(c) of
Ref.~\cite{Knispel}, with the theoretical and experimental spectra
independently normalized to their respective maxima.

The experimental spectrum exhibits a pronounced feature near
$-0.15$~eV, identified in Ref.~\cite{Knispel} as a Van Hove
singularity associated with the toroidal minimum of the Nb-derived
$d$ band around the $\Gamma$ point. Remarkably, the single-band
Wannier DOS places the corresponding maximum at approximately
$-0.16$~eV, without any empirical energy shift. This agreement
directly connects the isolated Nb-derived band of the low-energy
Hamiltonian to the characteristic energy scale observed by STS.

\subsection{Static Correlation Effects: DFT+U}

\begin{figure*}[t]
    \centering

    \includegraphics[
        width=0.9\textwidth
    ]{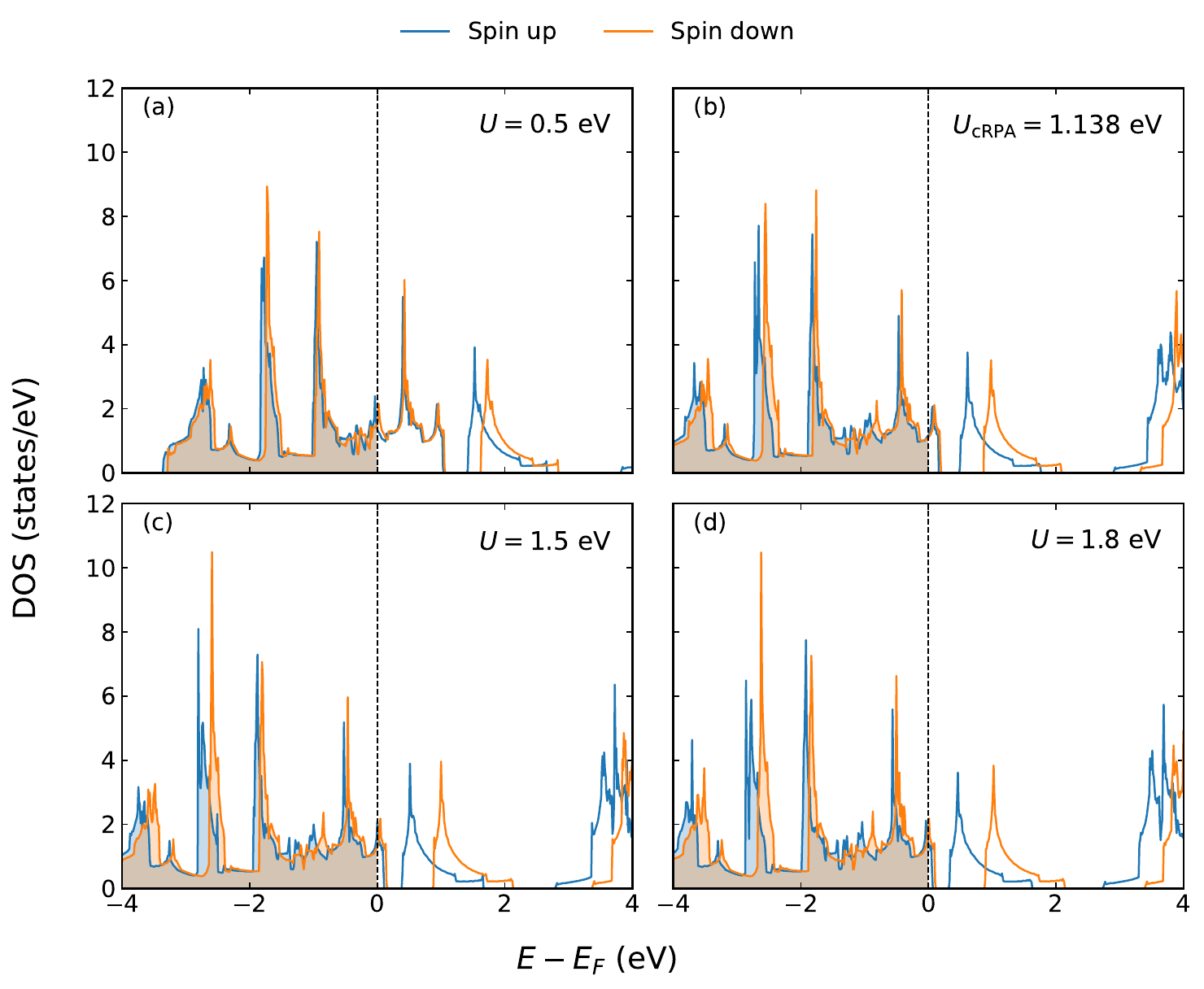}

    \caption{\justifying
    Spin-resolved density of states (DOS) of monolayer NbS$_2$
    obtained from DFT+$U$ for different values of the local
    Coulomb interaction:
    (a) $U=0.5$~eV,
    (b) $U_{\mathrm{cRPA}}=1.138$~eV,
    (c) $U=1.5$~eV, and
    (d) $U=1.8$~eV.
    The spin-up and spin-down contributions are shown separately.
    }

    \label{fig:dftu_spin_dos}
\end{figure*}

We first examine the effect of static local correlations within the DFT+\(U\) framework. As shown in Fig.~\ref{fig:dftu_spin_dos}, allowing spin polarization modifies the electronic structure, but does not lead to the opening of a gap over the interaction range considered. Instead, an increasing asymmetry between the spin-up and spin-down channels develops as \(U\) is increased. 

The most pronounced changes near the Fermi level occur in the spin-up channel, where spectral weight shifts toward lower energies. This redistribution is accompanied by a depletion of the density of states around the Fermi level and the development of a spin-polarized solution. Then, from the static DFT+\(U\) description, these results indicate that increasing the local Coulomb interaction favors spin polarization while preserving the metallic character of the system. The corresponding local magnetic moments are reported in Table~\ref{tab:magnetizacao_dftu}.

The results in Table~\ref{tab:magnetizacao_dftu} show a progressive
increase in the local magnetic moment with increasing Coulomb
interaction. The Nb moment increases from
$0.090\,\mu_{\mathrm{B}}$ at $U=0.5$~eV to
$0.162\,\mu_{\mathrm{B}}$ at
$U_{\mathrm{cRPA}}=1.138$~eV, reaching
$0.259\,\mu_{\mathrm{B}}$ at $U=1.8$~eV. The moments on the S atoms
remain small and antiparallel to the Nb moment across the
interaction range considered. Here,
the increasing Nb moment therefore reflects a growing tendency toward
spin polarization as the local interaction is strengthened, consistent
with the increasing asymmetry between the two spin channels observed in
Fig.~\ref{fig:dftu_spin_dos}. Although the development of spin polarization, no gap opens within
the interaction range considered. 

Moreover, the spin-polarized
DFT+$U$ solutions should not be interpreted as evidence of a specific
magnetic ground state, since DFT+$U$ treats the electronic interaction
at a static mean-field level \cite{Pavarini2021}. The subsequent DMFT calculations are
therefore restricted to the paramagnetic solution, allowing us to
isolate the effects of local dynamical correlations on the low-energy
spectral weight and quasiparticle properties.

\begin{table}[htbp]
\centering
\setlength{\tabcolsep}{10pt}       
\renewcommand{\arraystretch}{1.3}  
\begin{tabular}{cccc}
\hline
$U$ &
$m_{\mathrm{Nb}}$ &
$m_{\mathrm{S_1}}$ &
$m_{\mathrm{S_2}}$ \\

$(eV)$ 
&
($\mu_{\mathrm{B}}$) &
($\mu_{\mathrm{B}}$) &
($\mu_{\mathrm{B}}$) \\
\hline

0.5                         & 0.090 & -0.003 & -0.003 \\
\textbf{1.138}              & \textbf{0.162} & \textbf{-0.006} & \textbf{-0.006} \\
1.5                         & 0.221 & -0.007 & -0.007 \\
1.8                         & 0.259 & -0.008 & -0.008 \\

\hline
\end{tabular}

\caption{\footnotesize \justifying
Local magnetic moments obtained from spin-polarized DFT+$U$
calculations for different values of the local Coulomb interaction.
All values are given in Bohr magnetons ($\mu_{\mathrm{B}}$).}
\label{tab:magnetizacao_dftu}
\end{table}

\subsection{Correlation-Driven Evolution of the Spectral Function}
We can learn more about the low-energy single particle excitations
by examining the momentum-resolved spectral function
$A(\mathbf{k},\omega)$ in equation (\ref{eq:spectral_function}) for a given wavevector \textbf{k}.
In Fig~\ref{spectral_function} we present the momentum resolved DMFT
spectral function, $A(\mathbf{k},\omega)$, at $T\simeq290$~K for
increasing values of the local interaction $U$. The noninteracting
MLWF dispersion is superimposed for reference, allowing the
correlation-induced changes in the low-energy band to be followed
directly. ARPES measurements on NbS$_2$/Au(111) reveal a metallic
electronic structure, providing a useful experimental reference,
although substrate-induced hybridization and screening are not
included in the present model.

At $U=0.5$~eV [Fig.~\ref{spectral_function}(a)], the spectral
weight closely follows the MLWF dispersion, with a sharp band
crossing the Fermi level. Correlation-induced changes are therefore
small in this weak-coupling regime. At $U_{\mathrm{cRPA}}=1.138$~eV
[Fig.~\ref{spectral_function}(b)], the low-energy band remains
visible across the Brillouin zone but becomes noticeably broader
and more strongly renormalized. Despite this loss of coherence,
spectral weight continues to cross the Fermi level. The direct
single-band cRPA interaction therefore places the system in a
correlated metallic regime at this temperature. 
Further increasing the interaction to $U=1.5$~eV
[Fig.~\ref{spectral_function}(c)] strongly suppresses the
coherent band near $E_F$. The remaining low-energy spectral weight
becomes weak and diffuse, while spectral intensity is transferred
away from the Fermi level, indicating the onset of a
pseudogap-like regime. At $U=1.8$~eV [Fig.~\ref{spectral_function}(d)], the coherent
band crossing $E_F$ is no longer present. Instead, the spectral
weight is concentrated in broad features away from the Fermi level,
giving rise to a Mott-like spectrum within the present single-band
local-interaction model. Thus, within the present single-band DFT+DMFT description,
$U=1.8$~eV produces a Mott-like regime and does not qualitatively
reproduce the metallic behavior observed experimentally for
NbS$_2$/Au(111).
In contrast, the directly calculated single-band cRPA value,
$U_{\mathrm{cRPA}}=1.138$~eV, preserves a correlated metallic state
at the temperature investigated.

\begin{widetext}
\begin{center}

    \includegraphics[
        width=0.9\textwidth
    ]{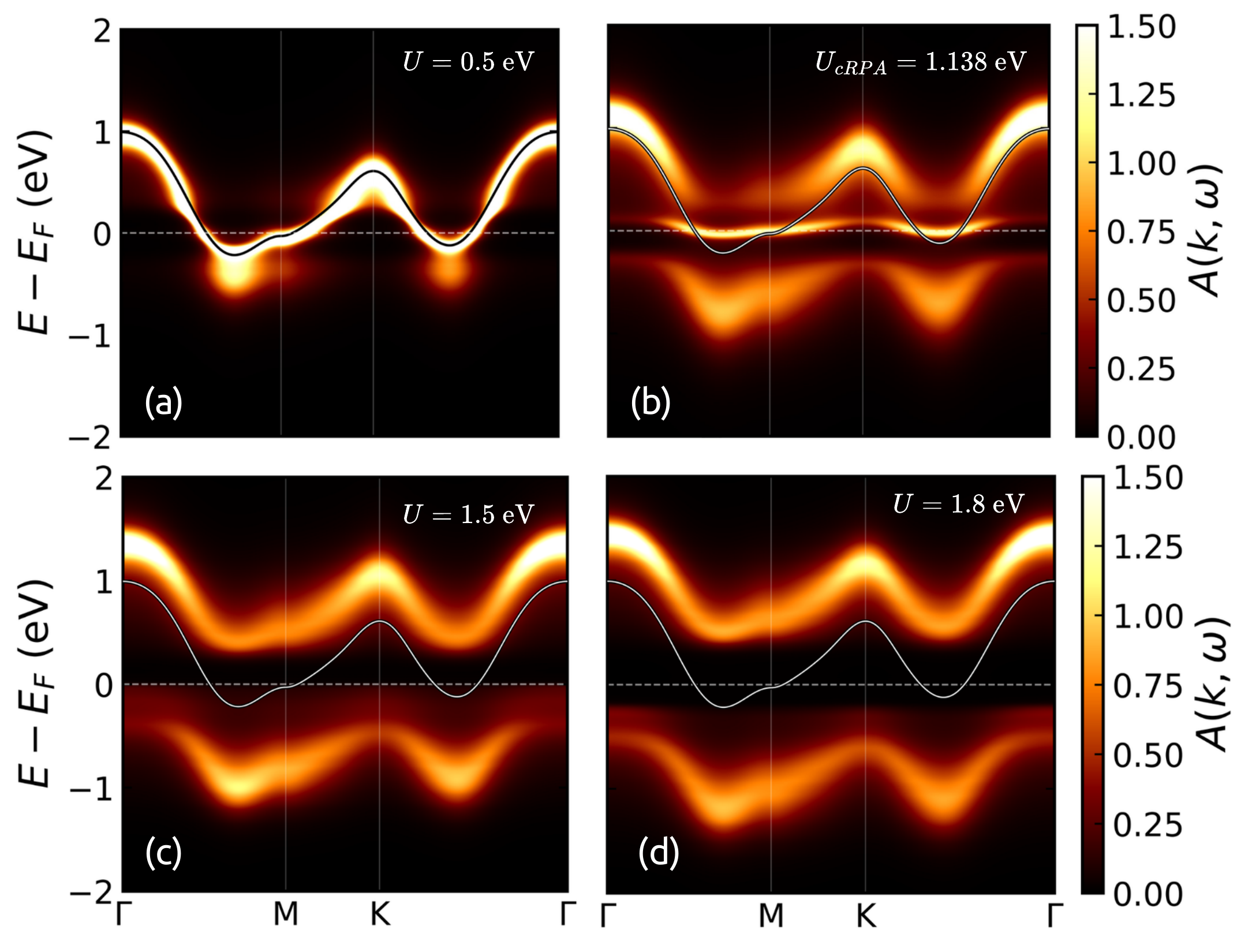}

    \begin{minipage}{1.0\textwidth}
    \footnotesize

    \refstepcounter{figure}
    \label{spectral_function}

    \textbf{FIG.~\thefigure.}
    \justifying
    DFT+DMFT momentum-resolved spectral function $A(\mathbf{k},\omega)$
at $T \simeq 290$~K for different values of the local Coulomb
interaction $U$: (a) $U=0.5$~eV,
(b) $U_{\mathrm{cRPA}}=1.138$~eV,
(c) $U=1.5$~eV, and
(d) $U=1.8$~eV.
The vertical axis represents $E-E_F$ (in eV), while the horizontal
axis corresponds to the high-symmetry path
$\Gamma$--M--K--$\Gamma$ of the Brillouin zone.
The MLWF band (solid line) is superimposed as a reference to the
non-interacting electronic structure.

    \end{minipage}

\end{center}
\end{widetext}

\subsection{Local Density of States (LDOS)}

In this section, we examine how the local density of states
$\rho_{\mathrm{loc}}(\omega)$, calculated from DFT+DMFT approch, evolves with
increasing local Coulomb interaction $U$ and describes the low-energy
spectral features of monolayer $1H$-NbS$_2$.

Figure~\ref{local_density} shows the DMFT local density of states
(LDOS) together with the noninteracting Wannier DOS. Consistent with
the momentum resolved spectra in Fig.~\ref{spectral_function},
increasing $U$ progressively redistributes spectral weight away from
the Fermi level. The low-energy weight remains substantial for
$U=0.5$~eV and $U_{\mathrm{cRPA}}=1.138$~eV, becomes strongly
suppressed at $U=1.5$~eV, and is nearly depleted at $U=1.8$~eV.
Thus, the LDOS provides the local counterpart of the evolution
identified in the spectral function.

Of particular interest is the behavior at the directly calculated
cRPA interaction, $U_{\mathrm{cRPA}}=1.138$~eV
[Fig.~\ref{local_density}(b)]. Finite spectral weight at $E_F$ is
retained at both $T\simeq290$~K and $T\simeq116$~K, showing that
the metallic solution persists upon cooling.
At the lower temperature, the low-energy spectral features become
sharper and more pronounced, while finite spectral weight is
preserved at the Fermi level. The temperature
dependence therefore reflects an increased coherence of the
low-energy excitations rather than a qualitative change of the
electronic state.

\begin{widetext}
\begin{center}

    \includegraphics[
        width=0.85\textwidth
    ]{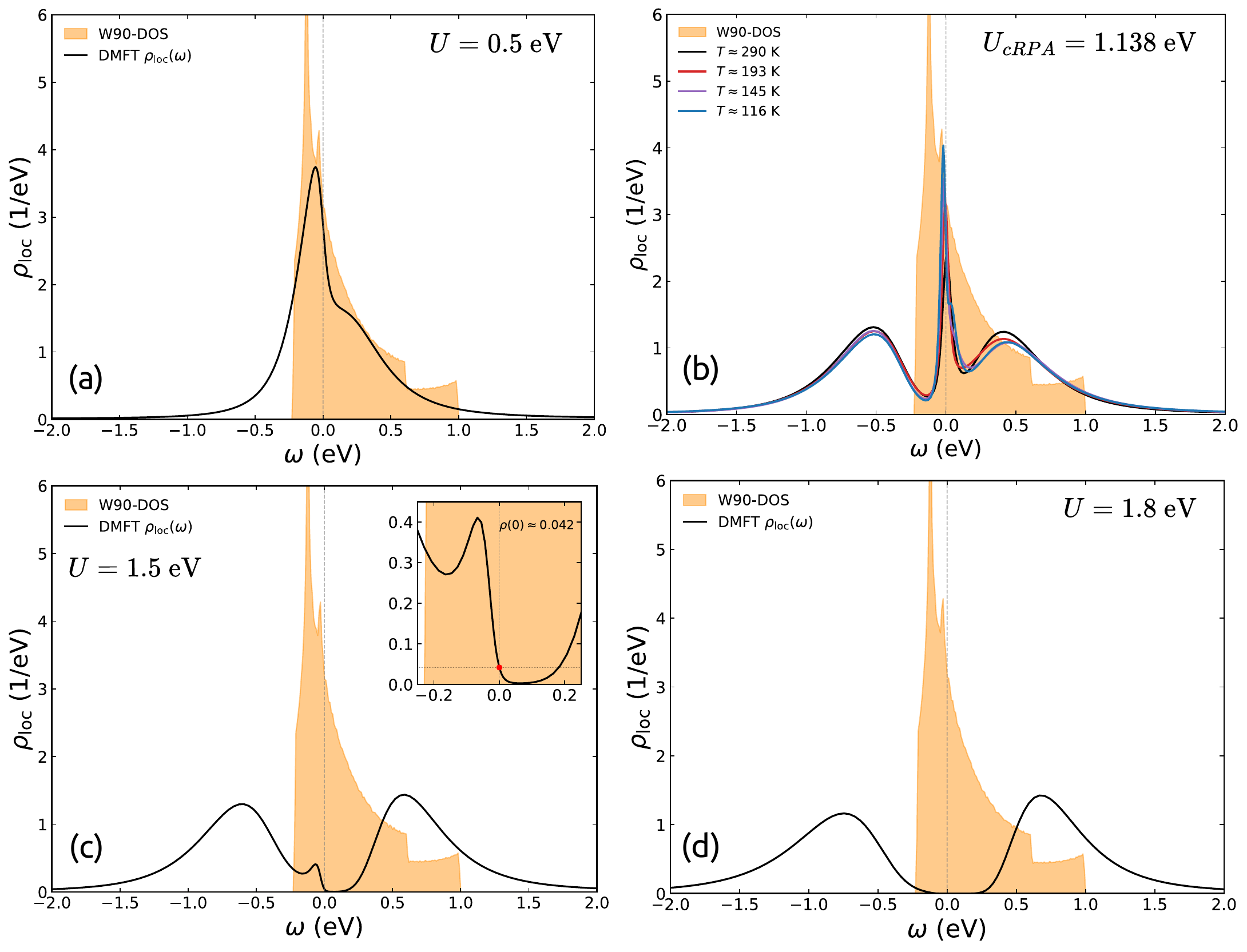}

    \begin{minipage}{1.0\textwidth}
    \footnotesize

    \refstepcounter{figure}
    \label{local_density}

    \textbf{FIG.~\thefigure.}
    \justifying Density of states obtained from DFT+DMFT for different values of the
local Coulomb interaction $U$: (a) $U=0.5$~eV,
(b) $U_{\mathrm{cRPA}}=1.138$~eV,
(c) $U=1.5$~eV, and
(d) $U=1.8$~eV.
The vertical axis represents the local density of states
(in states/eV), while the horizontal axis represents the energy
$E-E_F$ (in eV).
The calculations in panels (a), (c), and (d) were performed at
$T\simeq290$~K.
For the cRPA interaction, $U_{\mathrm{cRPA}}=1.138$~eV,
panel (b) shows results at
$T\simeq290$, $193$, $145$, and $116$~K,
corresponding to $\beta=40$, $60$, $80$, and $100$~eV$^{-1}$,
respectively.
The noninteracting Wannier DOS is shown as the filled orange area.
\end{minipage}

\end{center}
\end{widetext}

\subsection{Self-Energy, Quasiparticle Weight, and Effective Mass}

The local electronic self-energy provides a direct measure of the many-body renormalization introduced by the Hubbard interaction within DMFT. In the present calculations, the quantum impurity problem is solved using the continuous-time quantum Monte Carlo (CT-HYB) method, which naturally yields the self-energy on the imaginary-frequency axis. Access to real-frequency quantities requires analytic continuation of the corresponding Matsubara data, an intrinsically ill-conditioned procedure that may introduce additional uncertainties and numerical artifacts \cite{Jarrell1996}.

For this reason, we complement the real-frequency spectral analysis with quantities that can be determined directly from the low-frequency behavior of the Matsubara self-energy. In particular, we examine $\mathrm{Im},\Sigma(i\nu_n)$ and use its low-frequency slope to estimate the quasiparticle weight $Z$ and the corresponding effective-mass enhancement $m^*/m$. These quantities provide a continuation-independent characterization of the progressive quasiparticle renormalization and loss of electronic coherence as the interaction strength is increased.

Figure~\ref{fig:self_energy_Z_mass}(a) shows the imaginary part of the
self-energy for increasing local Coulomb interaction. As $U$ increases,
the magnitude of $-\mathrm{Im}\,\Sigma(i\nu_n)$ is progressively
enhanced at low Matsubara frequencies, indicating stronger many-body
scattering and a gradual loss of quasiparticle coherence. The change
is particularly pronounced between the weak-coupling case,
$U=0.5$~eV, and the cRPA interaction,
$U_{\mathrm{cRPA}}=1.138$~eV, while substantially larger
low-frequency values are reached for $U=1.5$ and $1.8$~eV.

This evolution is quantified in
Fig.~\ref{fig:self_energy_Z_mass}(b) through the quasiparticle weight
$Z$ and the corresponding effective-mass estimate
$m^*/m \approx 1/Z$. At $U=0.5$~eV, the quasiparticle weight remains
relatively large, $Z\approx0.705$. For the cRPA interaction,
$U_{\mathrm{cRPA}}=1.138$~eV, it is reduced to $Z=0.161$ at
$T\simeq290$~K, corresponding to $m^*/m\approx6.2$. Upon lowering
the temperature to $T\simeq116$~K, only a small change is observed,
with $Z=0.170$ and $m^*/m\approx5.9$. 
The weak temperature dependence of $Z$, together with its substantial
reduction from the weak interactions value, indicates strong
quasiparticle renormalization at the cRPA value.

For larger interactions, $Z$ decreases sharply to $0.035$ at
$U=1.5$~eV and $0.012$ at $U=1.8$~eV, reflecting the progressive
loss of quasiparticle coherence. At such small values of $Z$,
however, interpreting $1/Z$ as a conventional quasiparticle
effective mass becomes less meaningful because well-defined
quasiparticle excitations are no longer present. In this regime,
the strong low-frequency self-energy and the suppression of $Z$
are better regarded as signatures of the approach to the
incoherent, Mott-like regime identified in the spectral functions.

\begin{widetext}
\begin{center}

    \includegraphics[
        width=0.9\textwidth
    ]{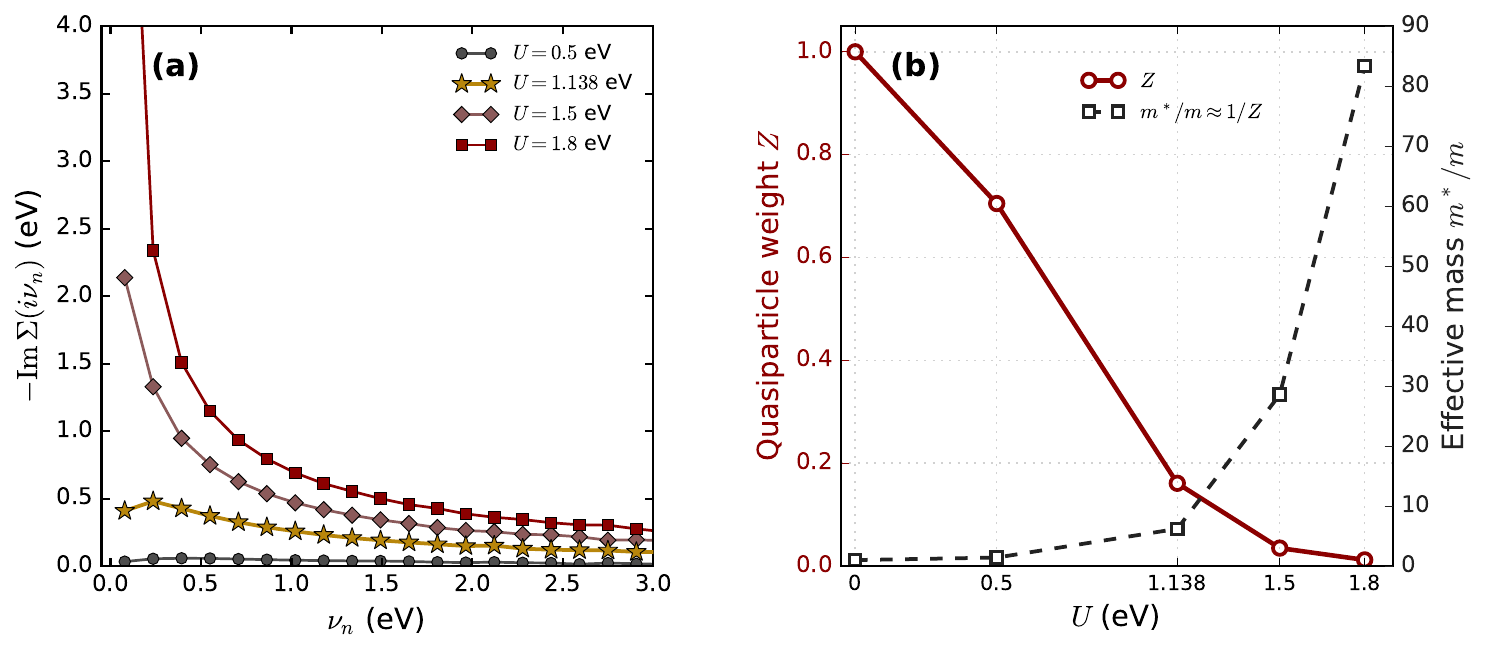}

    \begin{minipage}{1.0\textwidth}
    \footnotesize

    \refstepcounter{figure}
    \label{fig:self_energy_Z_mass}

    \textbf{FIG.~\thefigure.}
    \justifying
    (a) Imaginary part of the DMFT self-energy on the Matsubara axis,
$-\mathrm{Im}\,\Sigma(i\nu_n)$ (in eV), as a function of the
Matsubara frequency $\nu_n$ (in eV) for different interaction
strengths $U$.
(b) Quasiparticle weight $Z$ (left vertical axis) and effective-mass
enhancement $m^*/m \approx 1/Z$ (right vertical axis) as a function
of $U$ (in eV). All DMFT calculations shown in both panels were performed at
$T \simeq 290$~K.
    \end{minipage}

\end{center}
\end{widetext}

\section{Conclusion}

In conclusion, we constructed a single-band low-energy model for
monolayer $1H$-NbS$_2$ and determined the local Coulomb interaction
directly within the isolated Nb-$d_{z^2}$-derived target band.
The resulting static cRPA interaction,
$U_{\mathrm{cRPA}}=1.138$~eV, is therefore defined in the same
low-energy subspace subsequently treated within DMFT.

At $T\simeq290$~K, this interaction yields a strongly renormalized
metallic state with finite spectral weight at the Fermi level and a
quasiparticle weight $Z=0.161$. Upon lowering the temperature to
$T\simeq116$~K, the metallic spectral weight is preserved and
$Z$ changes only slightly to $0.170$, indicating that the correlation
regime remains qualitatively unchanged over the temperature range
considered. Independently, the single-band Wannier model reproduces
the characteristic low-energy Van Hove singularity observed by STS
near $-0.15$~eV, providing an experimental benchmark for the
downfolded Nb-derived band.

The direct single-band cRPA interaction is substantially smaller than
the effective value of approximately $U=1.8$~eV obtained in
Ref.~\cite{Loon} from a multiorbital construction. When used as the
only local interaction in the present single-band model, $U=1.8$~eV
drives the DMFT solution into a Mott-like regime, whereas
$U_{\mathrm{cRPA}}=1.138$~eV preserves a correlated metal. This
difference should not be interpreted as a discrepancy between two
estimates of a universal Hubbard parameter. The interaction in
Ref.~\cite{Loon} originates from a different correlated subspace and
was employed in an extended model including nonlocal Coulomb and
electron--phonon interactions. Our results instead emphasize that
the effective interaction and the many-body Hamiltonian in which it
is used must be defined consistently with the same low-energy
subspace \cite{carta2026,Merkel2024}.

\begin{acknowledgments}

The authors acknowledge the financial support from the Brazilian agencies FAPEMAT ( PQMT 2026 Scholarship Project), CNPq and CAPES.
The authors also express their gratitude to the National Laboratory for Scientific Computing for providing resources through the Santos Dumont supercomputer, and to the ``Centro Nacional de Processamento de Alto Desempenho em S\~ao Paulo'' (CENAPAD-SP, UNICAMP/FINEP - MCTI project) for support related to projects 897 and 909. Additional resources were provided by Lobo Carneiro HPC (NACAD) at the Federal University of Rio de Janeiro (UFRJ) for project 133. A.C.D. acknowledges financial support from FAP-DF grants 00193-00001817/2023-43 and 00193-00002073/2023-84, CNPq grants 408144/2022-0, 305174/2023-1, 444069/2024-0, and 444431/2024-1. L.A.R.J. acknowledges funding from FAPDF-PRONEM (grant 00193.00001247/2021-20),  CNPq (grants 301577/2025-0 and 444111/2024-7), and the Research Council of Norway (project 104795100). A.C.D. and L.A.R.J. also acknowledge funding from PDPG-FAPDF-CAPES Centro-Oeste grant number 00193-00000867/2024-94.  The authors gratefully acknowledge CAPES for the financial support provided for the publication of this work under the CC BY Open Access license through the Elsevier-CAPES agreement.

\end{acknowledgments}


\end{document}